\documentclass[sigconf]{acmart}

\renewcommand\footnotetextcopyrightpermission[1]{} 

\AtBeginDocument{%
  \providecommand\BibTeX{{%
    \normalfont B\kern-0.5em{\scshape i\kern-0.25em b}\kern-0.8em\TeX}}}

\usepackage{listings}
\usepackage{svg}

\colorlet{punct}{red!60!black}
\definecolor{background}{HTML}{EEEEEE}
\definecolor{delim}{RGB}{20,105,176}
\colorlet{numb}{magenta!60!black}

\lstdefinelanguage{json}{
    basicstyle=\normalfont\ttfamily,
    numbers=left,
    numberstyle=\scriptsize,
    stepnumber=1,
    numbersep=8pt,
    showstringspaces=false,
    breaklines=true,
    frame=lines,
    backgroundcolor=\color{background},
    literate=
     *{0}{{{\color{numb}0}}}{1}
      {1}{{{\color{numb}1}}}{1}
      {2}{{{\color{numb}2}}}{1}
      {3}{{{\color{numb}3}}}{1}
      {4}{{{\color{numb}4}}}{1}
      {5}{{{\color{numb}5}}}{1}
      {6}{{{\color{numb}6}}}{1}
      {7}{{{\color{numb}7}}}{1}
      {8}{{{\color{numb}8}}}{1}
      {9}{{{\color{numb}9}}}{1}
      {:}{{{\color{punct}{:}}}}{1}
      {,}{{{\color{punct}{,}}}}{1}
      {\{}{{{\color{delim}{\{}}}}{1}
      {\}}{{{\color{delim}{\}}}}}{1}
      {[}{{{\color{delim}{[}}}}{1}
      {]}{{{\color{delim}{]}}}}{1},
}

\definecolor{darkpastelgreen}{rgb}{0.01, 0.75, 0.24}

\newtheorem{definition}{Definition}

\begin{document}

\title{LongEval-Retrieval: French-English Dynamic Test Collection for Continuous Web Search Evaluation}


\author{Petra Galu\v{s}\v{c}\'{a}kov\'{a}}
\authornote{Both authors contributed equally to this research.}
\affiliation{%
  \institution{Univ. Grenoble Alpes, CNRS, Grenoble INP$^\dagger$\authornote{Institute of Engineering Univ. Grenoble Alpes.}, LIG}
  \city{38000 Grenoble}
  \country{France}
}
\email{petra.galuscakova@univ-grenoble-alpes.fr}

\author{Romain Deveaud}
\authornotemark[1]
\affiliation{%
  \institution{Qwant}
  \city{Paris}
  \country{France}}
\email{r.deveaud@qwant.com}

\author{Gabriela Gonzalez-Saez}
\affiliation{%
  \institution{Univ. Grenoble Alpes, CNRS, Grenoble INP$^\dagger$, LIG}
  \city{38000 Grenoble}
  \country{France}
}

\author{Philippe Mulhem}
\affiliation{%
  \institution{Univ. Grenoble Alpes, CNRS, Grenoble INP$^\dagger$, LIG}
  \city{38000 Grenoble}
  \country{France}
}
\email{philippe.mulhem@imag.fr}

\author{Lorraine Goeuriot}
\affiliation{%
   \institution{Univ. Grenoble Alpes, CNRS, Grenoble INP$^\dagger$, LIG}
  \city{38000 Grenoble}
  \country{France}
}

\author{Florina Piroi}
\affiliation{%
  \institution{Research Studios Austria, \\Data Science Studio}
  \city{Vienna}
  \country{Austria}
}

\author{Martin Popel}
\affiliation{%
  \institution{Charles University}
  \city{Prague}
  \country{Czechia}
}

\pagestyle{plain} 

\renewcommand{\shortauthors}{Galu\v{s}\v{c}\'{a}kov\'{a} and Deveaud, et al.}

\begin{abstract}

LongEval-Retrieval is a Web document retrieval benchmark that focuses on continuous retrieval evaluation. This test collection is intended to be used to study the temporal persistence of Information Retrieval systems and will be used as the test collection in the Longitudinal Evaluation of Model Performance Track (LongEval)\footnote{\url{https://clef-longeval.github.io/}} at CLEF 2023.
This benchmark simulates an evolving information system environment -- such as the one a Web search engine operates in -- where the document collection, the query distribution, and relevance all move continuously, while following the Cranfield paradigm for offline evaluation.
To do that, we introduce the concept of a {\it dynamic test collection} that is composed of successive sub-collections each representing the state of an information system at a given time step.
In LongEval-Retrieval, each sub-collection contains a set of queries, documents, and soft relevance assessments built from click models.
The data comes from Qwant, a privacy-preserving Web search engine that primarily focuses on the French market.
LongEval-Retrieval also provides a ``mirror'' collection: it is initially constructed in the French language to benefit from the majority of Qwant's traffic, before being translated to English.
This paper presents the creation process of LongEval-Retrieval and provides baseline runs and analysis.

\end{abstract}



\maketitle

\section{Introduction}

The LongEval test collection 
is proposed to support the evaluation of commercial and open-source state-of-the-art Web information retrieval systems. This test collection is dedicated to provide a large scale evaluation and it is able to cope with the temporal evolution of real Web data.
In order to conduct a proper evaluation of the search engines, we rely on the classical \textit{Cranfield paradigm}. Instead of providing one test collection, we build a succession of test collections, each of them composed of: a set of documents from Qwant's actual index, a set of topics acquired from actual user's queries, and two sources of relevance assessments that estimate which documents are relevant for each topic.

More specifically, the LongEval Web search test collection aims at evaluating IR systems with a focus on:
\begin{itemize}
\item{
the methods for connecting the predictivity and robustness of the offline and online evaluation, and}
\item{
the temporal persistence of IR systems.}
\end{itemize}

The corpus provides relatively large amounts of collected, filtered and cleaned Web pages (several millions) in two languages (original French documents and their automatic English translations). We consider the collection to be a useful experimental resource for Websearch engines and Information Retrieval systems in general. Moreover, the provided high-quality translations of both queries and documents might be helpful for the researchers working on cross-language retrieval.

\subsection{Use Cases}

The Longeval Websearch collection relies on a large set of data (corpus of pages, queries, user interaction) provided by a commercial search engine (Qwant). 
It is designed to reflect the changes of the Web across time, by providing evolving document and topic sets. 
To our knowledge, such features do not exist in a current test collections, at the scale we are focusing on.

The collection, as defined, aims at answering fundamental questions on the robustness and the stability of Web search engines against the evolution of the data.
Regarding Websearch evaluation, LongEval focuses on the following questions:
\begin{itemize}
    \item How does search engine behave as the collection evolves? Such question is especially important for commercial systems, as the satisfaction of users is the core of such systems.     
    \item When do we need to update an IR system as the collection evolves? If we are able to assess the decrease (if any) of a system on an evolving collection, we may then decide if the system needs to be updated.
\end{itemize}

 More globally, as the collection is large (millions of documents, thousands of queries, large assessments sets), it is usable as a good training source for deep learning methods. Such real source of data is rare enough to be very valuable for the community. Though our main focus is Web search, answering similar question on a robustness of the predictions is also in the interest of neural information retrieval systems trained and tested on different datasets.
 The collection, as defined, 
 can also help studying in detail the robustness of such deep models against novelty (documents, queries). Such data is a must for the community as this question is still largely open for IR.

\subsection{Related Work}
Some of the collections, such as Robust~\cite{voorhees2006trec} and Twitter~\cite{twitter:2017} contain information about the time at which was each document in the collection created. Such information might be used to study the changes in the collection and the behaviour of the search engines under these changes. However, collections which somehow consider an evolving environment as such are rare and, to our best knowledge, the only collection focused on the similar questions is the recent TREC-COVID dataset ~\cite{voorhees:2021}. 

TREC-COVID, proposes an evolving corpus, with topics from real questions extracted from forums. Compared to the LongEval collection, the number of documents and the number topics is rather small (few tens of thousands documents and 45 topics). TREC-COVID is focused on studying a  very specific topic, the COVID outbreak. In our case the topics are much broader, and the queries are actual user's ones. Thus, the LongEval collection is the only one large collection with up to date data (acquired in 2022), that exists to evaluate modern IR retrieval on evolving data. This also differentiates the LongEval collection from other cross-language collections which contain English-French language pair. French and English were for example part of the early TREC~\cite{Voorhees1999OverviewOT} and then CLEF~\cite{clef:2000,clef:2001,clef:2002} cross-language retrieval evaluation campaigns. From the later collections, French and English were in a limited way a part of CLEF eHealth Collections~\cite{kelly:2016,kelly:2019}. 

\begin{figure}[h]
    \centering
    \includegraphics[width=1\columnwidth]{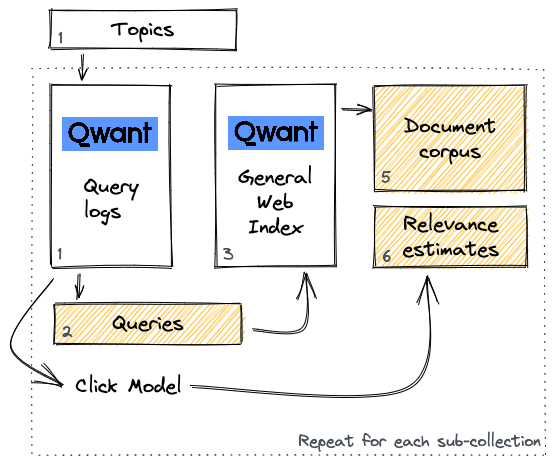}

    \caption{The general process for collecting and building the LongEval-Retrieval collection. Stroked boxes represent parts that are contained in each sub-collection.}

    \label{fig:qwant-general}
\end{figure}

\section{LongEval-Retrieval collection}
In this section, we describe the general acquisition process of the data from the Qwant Web search engine, and the creation of different components of the collection.

The overall acquisition is periodic and is recurrent over time in order to build a sequence of {\it sub-collections}. 
A sub-collection presents all the characteristics of a traditional test collection with queries, documents, and relevance judgments, except that it shares a common set of {\it topics} (which are not the queries themselves) with the other sub-collections. 

The overview of the acquisition is presented in Figure~\ref{fig:qwant-general}. It consists of:
\begin{enumerate}
    \item The acquisition of a set of \textbf{topics}, selected from the Web and social media. This acquisition is based on trending -- yet stable in the long-term -- topics and is performed only once for the entire LongEval-Retrieval collection. The selection of the set of topics is further described in Section~\ref{lab:topic-selection};
    \item The selection of \textbf{search queries} related to the topics, coming from the actual queries issued by users to Qwant. We detail the query selection method in Section~\ref{sec:sim-queries};
    \item  The creation of \textbf{relevance estimates}. We rely here on two ways of gathering such assessments: implicitly using Click Models~\cite{Chuklin2015} computed from Qwant query logs, and explicitly using manual assessments which would be gathered in the forthcoming future. Since each sub-collection can contain several thousands of queries, we will perform explicit assessments on a limited subset of manually selected queries. We present our methodology for generating relevance estimates in Section~\ref{sec:logbaseddata};
    \item The acquisition of the \textbf{document corpus} (Section \ref{sec:qwantcorpus}). This corpus from Qwant, is a union of: i) all the Web documents that have been displayed for each query of a sub-collection, and ii) a sizable random sample of the Qwant index. This protocol leads to a corpus that contains a mixture of relevant and non-relevant documents. The presented process handles the evolution of the Web pages, as the corpus is not only composed of URLs, but also of the Web page contents acquired at the time $t$.

\end{enumerate}

As described, the LongEval corpus evaluates systems along time. To do that, the acquisition presented is achieved periodically, typically each month. In each time period $t$, we create a collection composed by the queries, assessments and documents collected during this month. The full LongEval-Retrieval dataset, composed of a sequence of collections, is thus dynamically evolving. This allows us to create and provide test collections for different time periods.

\section{From Topics to Queries}
\label{sec:queriesdefinition}

As described above, the first step of the data acquisition is the definition of a set of \textit{topics}, which act as a proxy for controlling the themes expressed in the queries.
Those topics are later used for filtering and selecting the Qwant actual user's \textit{queries}.
The Qwant search engine processes approximately 200,000 queries per day. These queries follow the conventional long-tail distribution.
Therefore, we need to define a method for sampling intriguing and demanding queries that are likely to generate multiple user interactions.
We decide to select the query topics corresponding to trending topics (i.e., topics that are popular among the queries asked to Web search engines) in English and French.
This choice supports our use case of studying Web search engines, while it is also a good choice for examining temporal changes.

\begin{definition}
In the context of the LongEval-Retrieval test collection, a topic is a short multi-word term -- composed or either one or a few tokens -- with a broad and potentially ambiguous meaning.
\end{definition}

The topics then serve as an entry point for selecting a different set of queries for each sub-collection.
We allow the collection to have a shifting query set from one sub-collection to another in order to follow the potential drifts in the query distribution.

\begin{definition}
In the context of the LongEval-Retrieval test collection, a query is a multi-word chain of characters that is related to one or more topics.
\end{definition}

We detail how we selected the topics of LongEval-Retrieval in Section~\ref{lab:topic-selection}, before presenting in Section~\ref{sec:sim-queries} the method that we used to define a set of queries for each topic.

\subsection{Topics Selection}
\label{lab:topic-selection}

The list of topics has been generated once for the entire LongEval-Retrieval collection in May, 2022.
Topics were selected to ensure a balance between four elements: popularity, stability, generality, and diversity.
We targeted long-term trends of diverse domains to avoid selecting topics that were related to sudden interest shifts so as not to bias our collection towards events.
The set of topics is selected from English and French languages to ensure a multilingual validation of our findings, although a large majority of Qwant users are French.
To ensure some degree of stability in the entire dynamic collection, we defined a single set of topics that is common to all sub-collections.

\textbf{Topics should be popular}: this criterion should ensure that there will be enough potential queries covering this topic in Qwant's query log, and in turn, enough user feedback.

\textbf{Topics should be stable} in time: it is important to be able to consider topics that do not evolve over time, so that we can really assess longitudinally the behavior of systems.
To do that, we measure the persistence and recurrence in time on historical trending topics using Google Trends\footnote{\url{https://trends.google.com}} and the Qwant's query log.

\textbf{Topics should be general} enough to cover numerous queries. 
We sampled queries from a 1-month Qwant's query log, and kept the queries that overlapped with our candidate topics.
We considered topics for which a large number of queries matched ($\geq$ 1,000 queries) as general enough.

Table~\ref{tab:keyword} presents the set of topics that we used to further select the queries of the sub-collections.

\begin{table}[ht!]
\resizebox{\columnwidth}{!}{%
\begin{tabular}{@{}lll@{}}
\toprule
  & Topic & English Description \\ \midrule
1 & eau & water \\
2 & nourriture & food \\
3 & espace & space \\
4 & voiture & car \\
5 & argent & money \\
6 & manifestation & protest \\
7 & virus & virus \\
8 & terre & earth \\
9 & énergie & energy \\
10 & police & police \\
11 & loi & law \\
12 & travailleurs & workers \\
13 & guerre & war \\
14 & invasion & invasion \\
15 & président & president \\
16 & pétrole & oil \\
17 & impôts & taxes \\
18 & votants & voters \\
19 & fraude & fraud \\
20 & elisabeth borne & (French prime minister, May, 2022) \\
21 & changement climatique & climate change \\
22 & fête du travail & (French holiday, usually associated with protests) \\
23 & eurovision & eurovision \\
24 & jacques perrin & (French actor) \\ 
25 & regine & (French singer) \\
26 & heartstopper & - \\
27 & fete des meres & mother's day \\
28 & johnny depp & - \\
\bottomrule
\end{tabular}%
}
\caption{List of the French topics of the LongEval-Retrieval collection that were the basis for selecting the queries. The English description is informative and has not been further used to build the collection.}
\label{tab:keyword}
\end{table}

\subsection{Queries Selection}
\label{sec:sim-queries}

The topics previously extracted following the methods described in the previous section are a first step towards the identification of real user queries that are issued on Qwant, which we will use for each sub-collection.
Qwant answers a portion of user queries with their own technology while the rest is forwarded to a third party search engine.
In order to ensure that all displayed documents are contained in Qwant's index, we restricted the query distribution that we consider to the queries that Qwant actually answers.

We use a simple text processing technique to map general topics to queries that are answered by Qwant.
Let $\mathcal{Q}$ be the set of all queries that are answered by Qwant, and $\mathcal{T}$ be the set of topics we defined in Section~\ref{lab:topic-selection}.
For each topic $t \in \mathcal{T}$, we select all the queries $\mathcal{Q}_t$ from $\mathcal{Q}$ that contain $t$ as a sub-string (denoted by $\subseteq_{str}$):
$$
\mathcal{Q}_t = \{ q | q \in \mathcal{Q}, t \subseteq_{str} q \}
$$
Then, for the full set of topics, we have:
$$
\mathcal{Q}_\mathcal{T} = \cup_{t \in \mathcal{T}}{\mathcal{Q}}_t
$$

As this filtering can lead to several tens of thousands of queries per topic when considering several days/months of query logs, we applied a top-$k$ selection for each $\mathcal{Q}_t$, noted $\mathcal{Q}^{top_k}_t$, using which we only keep the $k$ most frequently asked queries on Qwant for each topic. 
Finally, $\mathcal{Q}^{top_k}_\mathcal{T}$ denotes union of $\mathcal{Q}^{top_k}_t$ over the topics.

\subsection{Queries Filtering}
Following the initial query generation process, the queries undergo automatic filtering to select those with at least 10 relevance assessments (as detailed in the following section)\footnote{It should be noted that some queries may have fewer than 10 assessments in the final collection due to further corpus filtering applied after this step}. Subsequently, a manual check is performed by a human annotator to ensure query quality. During this process, queries with similar objectives, such as `achat voiture' and `acheter voiture', `anti virus' and `antivirus', or `bareme impots' and `barême impots', are merged. Queries referring to adult content are removed from the collection.

\section{Relevance estimates}
\label{sec:logbaseddata}

The relevance estimates for LongEval-Retrieval are obtained through automatic collection of user implicit feedback. However, in Section~\ref{sec:discussion}, we discuss the future completion of these estimates with explicit relevance assessments.

In order to be consistent with the creation of the query and document collections, we rely on user clicks as an implicit feedback to automatically infer the relevance of different documents.
 Qwant is committed to protecting online privacy, and does not track information such as multiple clicks, dwell times, or query reformulations. The only information logged by Qwant is the query, the displayed documents, and the rank at which a click occurred.

Raw clicks cannot be used as a signal of relevance due to their tendency to be noisy and heavily biased~\cite{joachims-2005} towards the top-ranked results.
Noise comes from the fact that a click does not necessarily indicate relevance, while a lack of click does not indicate irrelevance. 
Aggregating larger samples of query logs can remove noise, but statistical biases remain.
Such biases can result from various factors, such as position bias towards top-ranked results, presentation bias towards visually appealing results, or trust bias towards results from familiar domains~\cite{Oosterhuis-2021}.

\subsection{Click Models}
To tackle this problem, we opted to {\it debias} the click data by estimating Click Models.
A Click Model~\cite{zhang2011user,chuklin2013click} is the base to infer the user relevance of a document from search log data. 
It computes the estimates of a document attractiveness given a query and thus it well tackles the problem of being able to use the users' interaction while avoiding sharing private data and reducing noise and bias.
Many click models have been developed over the years with the goal of better modelling the clicking behaviour of users, mostly by investigating sessions of multiple clicks for a given query.
In our case, since Qwant is a privacy-preserving search engine and does not track any search session, we cannot rely on the more advanced Click Models that consider multi-query and multi-click sessions.

We thus implemented the simplified version of the Dynamic Bayesian Networks (DBN)~\cite{Chapelle2009-rg} with a session length of 1, which comes down to the original Cascade Model~\cite{Craswell2008-pa}.
The Cascade Model assumes the following user model: users scan a Search Engine Results Page (SERP) from top to bottom, skipping non-attractive documents and clicking on what they believe will be relevant based on attractiveness.
By fitting such a model, we were able to use the learned attractiveness parameter $\alpha$ as a soft relevance assessment.
Such assessment of the relevance of a document given a query is both a) noise-free and b) unbiased with respect to the underlying user model.
The $\alpha$ parameter of the model is then equal to the probability that a document would be attractive given a query. 
Maximum Likelihood Estimation (MLE) can be used to estimate $\alpha$ for each pair of query-document.
For a given query $q$ and a document $d$, the attractiveness probability is estimated as follows:
$$
\alpha_{q,d} = \frac{1}{|\mathcal{S}_{q,d}|}\sum_{s \in \mathcal{S}_{q,d}} c_s(d)
$$
where $\mathcal{S}_{q,d}$ is the set of all entries in Qwant's query log where document $d$ has been displayed in a SERP above or at the rank of the clicked document, and $c_s(d)$ is a binary variable indicating where document $d$ has been clicked in the entry $s$.

We estimated this parameter by using MLE over Qwant's query log filtered for the queries and the time period of each sub-collection.
The attractiveness $\alpha$ then provides us with a surrogate for document relevance for each query.
However, traditional IR metrics such as Precision, Mean Average Precision, or nDCG rely on integer relevance assessments (either binary or scaled). 
This makes 
the attractiveness probabilities as relevance estimates impractical for computing IR metrics.
We thus transformed $\alpha$ into discrete values by reversing a function $f(g)$~\cite{chapelle2009expected} used for mapping a discrete relevance judgment $g$ to a probability of relevance, based on the Discounted Cumulative Gain discount function:
$$
f(g) = \frac{2^g -1}{2^{g_{max}}}
$$
where $g_{max}$ is the maximum grade of relevance (i.e. $g_{max} = 1$ for binary relevance judgments).
We simply reversed the function as follows:
\begin{align*}
& P =  \frac{2^g -1}{2^{g_{max}}} \\
 \Leftrightarrow & P \times 2^{g_{max}} = 2^g -1 \\
  \Leftrightarrow &  g = log_2(P \times 2^{g_{max}} + 1)    
\end{align*}
where $P = \alpha_{q,d}$ in our case.

Using Qwant's clicklog, we then compute both probability-based and discrete relevance estimates for all query-document pairs that we gathered using the methods previously depicted.
We specifically used $g_{max} = 2$, which corresponds to:
\begin{itemize}
    \item 0 = not relevant,
    \item 1 = relevant,
    \item 2 = highly relevant.
\end{itemize}

\subsection{Post-filtering}
The last step of building the relevance estimates is their post-processing according the filtering done on the queries. As described above, the queries were manually checked and some of the queries were removed. In such case, all the relevance estimates corresponding to such queries were removed as well. Similarly, some of the queries were merged and in such case, the corresponding estimates were also merged. If, during the merge, the assessments of the queries were having contradictory values (e.g. the document was relevant according to one query and not relevant according to the second query), these relevance estimates were excluded from the collection. 

\section{Document Corpus}
\label{sec:qwantcorpus}

The final part of the test collection is related to the corpus of documents, on which the relevant documents are selected to be retrieved for a specific query. The first step for creating the document collection is to extract from the index the content of all the documents that have been displayed in SERPs for the queries that we selected (see Section~\ref{sec:sim-queries}). In addition to these documents, potentially non-relevant documents are randomly sampled from Qwant index in order to better represent the nature of a Web test collection.

\subsection{Random Sampling}
Two major problems that could occur if we built the document collection only using the documents of the relevance assessments: 1) the collection would be heavily biased towards Qwant's ranking function, and 2) the prevalence of relevant documents would be artificially high.
We alleviated this problem by sampling random documents for each original topic and introducing them into each sub-collection.
For each topic $t \in \mathcal{T}$, we randomly selected at most $n = 100,000$ documents among those that matched the word tokens of $t$\footnote{We used a very basic AND matching, hence selected documents had to contain all the word tokens of the topic.}; this number could be lower depending on the number of documents actually matching the topic.
Such selection avoids oversimplifying the corpus and the search task, as these documents are not completely randomly picked from Qwant index.

\subsection{Cleaning Process}
The collection does not only provide the url of each document in the corpus, but also cleaned versions of these documents. To do so, we first extract the text content from the websites, for which we use the internal Qwant implementations. Thus, we are able to use exactly the same representations of documents as Qwant uses for ranking the documents. 
We also apply an adult and spam content filter on the built collection. Even though this filtering is quite strict, adult and spam content is still very frequent in the collection. 

\section{English Translations}
\label{sec:multilingual}
Given that the vast majority of Qwant users are French speakers, the search engine's primary focus is on searching for and analyzing French queries and data. However, to make the collection more accessible to non-French speakers, we also provide English translations for both queries and documents. As a result, the collection can be used for monolingual English searches and is also well-suited for cross-lingual retrieval purposes.

Initially, all the queries and documents were created in French. To generate their English counterparts, we used a machine translation system to translate the French queries and documents into English. For the translation, we used the French-English CUBBITT (Charles University Block-Backtranslation-Improved Transformer Translation) system
\citep{popel-et-al-2018},
available at the LINDAT/CLARIAH-CZ infrastructure\footnote{
 \url{https://lindat.cz/services/translation}
}. 
Naturally, the quality of the translations of the queries, which are often very short, is much lower than the quality of the document translation. This is due the 
mismatch of domain of training data which are not suitable for query translation and also due to that the translation system has in the case of the documents much more contextual information available.

To translate the documents, we first segment all the websites into sentences using the French model implemented by Spacy\footnote{\url{https://spacy.io/}}. We then detect the language of each sentence using Fasttext~\cite{joulin2016bag} and only translate the sentences that are highly likely to be  French. Otherwise, we skip the translation and keep the original sentence. We also exclude single-character sentences from translation, as they often lead to catastrophic errors. To ensure tractability, we limit the translation to truncated sentences of up to 500 bytes. Our experiments (see Section~\ref{ssec:prelimtests}) demonstrate that this approach has little impact on the evaluation measures.

\section{LongEval Collections}
\label{sec:qwantdyn_testcoll}
LongEval is a shared task, which will be for the first time organized in 2023 at CLEF. The task is dedicated to evaluating the temporal persistence of Information Retrieval (IR) systems and Text Classifiers. It is motivated by recent research showing that the performance of the models drops as the test data becomes more distant, with respect to time, from the training data. This is true for classification~\cite{Lukes2019,florio2020time,alkhalifa2022building}, but also the research in information retrieval shows that deep neural network-based  IR systems are dependent on the consistency between the train and test data~\cite{https://doi.org/10.48550/arxiv.2204.12755}. To be able to study this, one needs several test collections created over sequential time periods, which would allow doing observations at different time stamps $t$, and most importantly, comparing the performance across different time stamps $t$ and $t'$. We want to specifically investigate two particular scenarios: short-term persistence and long-term persistence. The short-term persistence task aims to assess the performance difference between $t$ and $t'$ when $t'$ is occurs right after or shortly after $t$. In the long-term persistence task, we aim to examine the performance difference between two $t$ and $t''$, when $t''$ occurs several months after $t$ (and thus $|t''-t| > |t'-t|$).

So far, we described creation of the collection at the single time stamp $t$ referring to a single month. We repeat this process of collecting the data over several months and create the Train Collection $t$ and two test collections $t'$ and $t''$.

\subsection{Train Collection}
Train collection is collected during June 2022 and was already released on the Lindat infrastructure\footnote{\url{http://hdl.handle.net/11234/1-5010}}. The document corpus consist of 1,570,734 Web pages. As we would like to be able to compare the systems between $t$ and $t'$ or $t''$, we further split the queries in this Train collection into train and heldout queries. The collection consists of 672 train queries, with corresponding 9,656 assessments and 98 heldout queries with corresponding 1,420 assessments. There are thus in average 14 assessments per query. About 73\% of the assessments are non-relevant (7,030 assessments on the train queries in total), 21\% are relevant (2,028 assessments) and 6\% are highly relevant (598 assessments). 

\begin{table*}[htb]
\begin{tabular}{c|ll} 
\bottomrule
Query ID & French Query & English Query\\
\hline
q06229550 & bareme impots & Taxation \\
q06223863 & consommation eau & consumption water \\
q06221247 & gateau aux pommes & apple cake \\
q06225303 & offre emploi & offer of employment \\
q06226031 & terrence malik & terrence malik \\
q06226042 & tesla voiture & tesla car \\
q06223096 & visiter les chateaux de la loire & visit the castles of the loire \\
\toprule
\end{tabular}
\caption{Examples of the train queries.}
\label{tab:queries}
\end{table*}

\begin{table*}[htb]
\begin{tabular}{p{0.45\textwidth} p{0.45\textwidth}}
\bottomrule
Highly Relevant Document & Non-Relevant Document\\
\hline
Bordeaux Airport
-
Merignac
- Official website
Re-opening of the billi terminal: all the information you need for your next trip
Our billi terminal has reopened to operate some easyJet and Ryanair flights.
Shops were also reopened, in particular to offer a catering service to travellers.
This decision will make it possible to deploy on the entire infrastructure the material and human resources necessary for the smooth running of the journey.
Important: All other airlines: concourse A Flight boarding: check your concourse and boarding gate in real time on our website or on the terminal screens when you arrive.
With large crowds, think ahead of your visit:
Due to the heavy summer traffic, passengers are advised to anticipate their arrival well in advance.
Wings for the World News and events Published: 10 June 2022
&
Parking near Bordeaux Airport
Merignac
|
Beep Valet
Parking
contact@beep-valet-parking.com
BEEP
VALET
Choose your seat type
Parking
secure + Shuttle offered + Shuttle Airport
You are brought to the airport from the car park Reservation
Airport
Free Parking + Shuttle Parking Available
Your windscreen needs replacing?
We offer parking Reservation Parking Offered
Secure payment by bank CIC Benefits of your car park
Merignac
airport
RESERVE
RAPIDE
GAREZ
EASY TALK SEREIN Close
airport Mérignac
Our tarmac car park is located 3 km from Bordeaux-Mérignac airport.
Take 5 minutes of transfer time in our comfortable shuttle.
If needed, child boosters and baby seats are offered.
Airport Shuttle\\
\bottomrule
\end{tabular}
\caption{Examples of the translated documents relevant and non-relevant for the query ``aeroport bordeaux'', which was translated as ``airport''.}
\label{tab:documents}
\end{table*}

We display several randomly selected examples of the queries and their translations in Table~\ref{tab:queries}. Though the majority of the queries, as displayed, is translated reasonably, some of the translations still failed. For example the query ``cuisson gigot agneau'' (meaning ``cooking leg of lamb'') is translated as ``leg'' repeated numerous times. The original French queries are relatively short -- the most of the queries consists of either three (42\%) or two (30\%) words. Only a single query consist of 7 words and there is no longer query than that. In Table~\ref{tab:documents}\footnote{Documents are shortened for the display purposes.} we show an example of highly relevant and non-relevant document.

While we publish the assessments for the train queries, we keep the assessments for the heldout queries hidden from the participants and only release them after publishing the official task results. The participants are required to submit the results for either one or both test collections and also to submit the results on the heldout queries. This setup allows us to measure the differences of the submitted systems between $t$ and either $t'$ or $t''$.   

\subsection{Test Collections}
We further plan to release two test collections. The short-term persistence collection is collected over July 2022 and the long-term persistence collection over September 2022. The collections will be according to the shared task timeline released in April 2023.

For the evaluation, we plan to apply nDCG measure to all the systems submitted to the task on the heldout data and on the test sets. This classical evaluation measure should be well suitable for evaluating Web search, for which the discount emphasises the ordering of the top results. Apart from this, we plan to calculate Relative nDCG Drop (RnD) measured by computing the difference between nDCG on the heldout queries and test collection. This measure supports the evaluation of the impact of the data changes on the system's results. Using these two evaluation measures, a system that has good results using nDCG, and also good results according to the RnD measure is considered to be able to cope well with the evolution over time.

\section{Experiments}
\label{ssec:prelimtests}
We describe here preliminary tests that are conducted on the training corpus released, in a way to show that the data provided is hard enough to be a good source for research purposes, with a large margin of improvement for systems. It allows also to validate to some extend our automatic relevance assessment process. In this first step, we use two default IR systems, Terrier \cite{macdonald2012puppy} and Anserini \cite{anserini:2017}, on both French and English corpora, using their respective default BM25 parameters, with results sizes of 1000 documents. We apply French stopwords and stemming in Terrier 
when processing the French collection, but we do not use any specific French processing in Anserini.
We also choose these systems to check both of our data formats: Anserini relies on json format and Terrier on TREC format.  The collection contains the data in both these formats, what allows easy application of both Anserini and Terrier systems.

The results are presented in Table~\ref{tab:resultsBM25train} for the train query set and in Table~\ref{tab:resultsBM25heldout} for the heldout query set. The first two rows present the results on the French corpus: we see that both systems perform reasonably well, with P@10 close to 0.1 and NDCG around 0.32.  In both cases, the Recall@1000 is close to 0.75, which is a reasonable score supporting experimentation wih reranking IR approaches. The results of both BM25 implementations are, not surprisingly, very similar and the results for the English corpus are 10\% lower that the French corpus. 
The interesting point is that the translation does not impact dramatically the results: this validates the quality of the translations and confirms that the translated English corpus may be used with a confidence.
The comparison of the results on the train and heldout queries shows that all systems perform very similarly on these query sets. This confirms that the heldout queries are good representatives for the Train collection.

\begin{table}[htb]
\begin{tabular}{|c|c|c|c|c|c|} 
\bottomrule
System & Language & MAP & P@10 & NDCG & Recall\\
\hline
\hline
Anserini & fr & 0.1617	& 0.0987	& 0.3102	& 0.7368\\
Terrier & fr &	0.1767	& 0.1109	& 0.3308	& 0.7695\\
\hline
Anserini	& en	& 0.1372	& 0.0887	& 0.2768	& 0.6766\\
Terrier & en & 0.1460	& 0.0939	& 0.2897	& 0.7059\\
\toprule
\end{tabular}
\caption{BM25 results for Anserini and Terrier (top-1000 documents), on the Train collection and the \textit{train} query set of LongEval.}
\label{tab:resultsBM25train}
\end{table}

\begin{table}[htb]
\begin{tabular}{|c|c|c|c|c|c|} 
\bottomrule
System & Language & MAP & P@10 & NDCG & Recall\\
\hline
\hline
Anserini & fr & 0.1676 & 0.1112 & 0.3249 & 0.7573\\
Terrier & fr &	0.1904 & 0.1184 & 0.3540 & 0.8119\\
\hline
Anserini & en & 0.1267 & 0.0878 & 0.2646 & 0.6367 \\ 
Terrier & en & 0.1326 & 0.0908 & 0.2689 & 0.6372 \\
\toprule
\end{tabular}
\caption{BM25 results for Anserini and Terrier (top-1000 documents), on the Train collection and the \textit{heldout} query set of LongEval.}
\label{tab:resultsBM25heldout}
\end{table}


\section{Discussion}
\label{sec:discussion}
The average number of documents retrieved by the BM25 models described in the previous section for which we have assessments available is very small: in average there are only 2 or 3 assessed documents in the top 10 retrieved documents and only around 7 assessed documents in the top 100 retrieved documents. These numbers are similar for all the runs, but slightly higher for Terrier and especially when it is applied on the English data. We would like to deal with this problem by using the explicit relevance assessments from the users, using classical TREC pooling methods.

This explicit feedback will be provided for a selected subset of  queries defined in Section~\ref{sec:queriesdefinition} and for the pooled list of documents which will be retrieved by the systems submitted to the LongEval-Retrieval track. The relevance assessments are thus expected to be available in summer 2023. This step is especially important to be able to compare the predictions done by the click model with a more precise human assessments and thus, in turn, to compare offline and online evaluation methods. It might also help to 
make evaluation more fair for the systems utilizing English data, as these can be expected to retrieve different set of documents as the systems utilizing French data.

Moreover, the explicit assessments might help to deal with the noise in the implicit ones. Proposed test collection aims at representing a Web environment, with a focus on topics and documents which change over time. Given these characteristics of our test collections, we may have possible noise/error induced on the relevance assessment when we cluster several one-click user logs, in order to infer the complete user interaction. It is due to the fact that Qwant stores limited user interaction information.

Based on the recent observations from TREC~\cite{craswell:2021}, the selection of the queries for the pool will be crucial for the explicit assessments. While the selected queries need to be representative and also need to have enough relevant documents, the number of relevant documents still needs to be limited. If the amount of relevant documents is too high, it might not be possible to easily distinguish the quality of the systems, which we would like to compare.

\section{Conclusion}
This paper describes the LongEval-Retrieval test collection, which is designed specifically for evaluating information retrieval systems with evolving data. The collection is constructed from Qwant data, including Web documents and real user input. The described data collecting and processing methods are able to cope with large amount of data, by generating automatic queries and their relevance estimates from query logs. We create a corpus which integrates SERP results as well as ``background'' documents from the Qwant index. The corpus is initially in French, and LongEval-Retrieval also provides English translations of the documents and queries.

We describe in detail the process of creating the collection and also the published Train collection which uses this process. We also provide initial experiments which confirm usability of the collection and provide an initial baseline. In the following step, we plan to further provide two Test collections with similar characteristics. In the future, we also plan to integrate manual assessments into the collection, so that we can compare manual and automatic assessments in a large scale. LongEval-Retrieval is currently able to cope with queries evolution, according to a fixed set of initial topics, but in the future, we want to extend our proposal so that we also handle evolution of topics.

Regarding the initial objectives of LongEval-Retrieval, we aim to connect the predictivity and robustness of the offline and online evaluation approaches. Using the proposed evaluation measures, we will be able to explore the temporal persistence of IR systems across several corpora that evolve, compared to a reference one. By doing so, the collection will provide a more comprehensive and accurate evaluation of IR systems in real-world settings, where the data is constantly changing over time.

\section*{Acknowledgments}
This work is supported by the ANR Kodicare bi-lateral project, grant ANR-19-CE23-0029 of the French Agence Nationale de la Recherche, and by the Austrian Science Fund (FWF, grant I4471-N). This work has been using services provided by the LINDAT/CLARIAH-CZ Research Infrastructure (https://lindat.cz), supported by the Ministry of Education, Youth and Sports of the Czech Republic (Project No. LM2018101 and Project No. LM2023062) and has been also supported by the Ministry of Education, Youth and Sports of the Czech Republic, Project No. LM2018101 and Project No. LM2023062 LINDAT/CLARIAH-CZ. 

\bibliographystyle{ACM-Reference-Format}
\bibliography{references}
\end{document}